\documentstyle[12pt,epsf]{article}
\def\sp#1{{\rm Tr}\biggl( #1 \biggr)}
\def\norm{{v^2\over \Lambda^2}}

\def\ra{\rightarrow}
\def\prd#1#2#3{{\it Phys. Rev.} {\bf D#1} #2 (19#3)}
\def\pl#1#2#3{{\it Phys. Lett.} {\bf #1B} #2 (19#3)}
\def\np#1#2#3{{\it Nucl. Phys.} {\bf B#1} #2 (19#3)}

\begin{document}
\begin{titlepage}
\def\ba{\begin{array}}
\def\ea{\end{array}}
\def\thefootnote{\fnsymbol{footnote}}
\begin{flushright}
	ISU-HET-95-2\\
	March, 1995
\end{flushright}
\vspace{1in}
\begin{center}
{\large \bf EXPERIMENTAL SIGNATURES OF A \\PARITY VIOLATING
ANOMALOUS COUPLING $g_5^Z$}\\
\vspace{1in}
        {\bf   G.~Valencia}\\
        {\it   Department of Physics and Astronomy\\
               Iowa State University\\
               Ames, IA 50011}\\
\vspace{1in}
     %	{\large \bf ABSTRACT}
\end{center}
\begin{abstract}

I discuss the experimental signatures of a parity violating but $CP$
conserving interaction in the symmetry breaking sector of  the
electroweak theory.

\end{abstract}

\end{titlepage}

\section{Introduction}

The standard model of electroweak interactions has now
been tested thoroughly in a number of experiments. The
only sector that has not been tested directly is the
electro-weak symmetry breaking (or Higgs) sector. It is
very important to understand in detail the experimental
signatures for the symmetry breaking sector. These vary
from the direct search for new particles such as a Higgs
boson, to the search for indirect manifestations of the
existence of these new particles.

A convenient parameterization of  these indirect effects
of new particles at energies below threshold for their
production is that of anomalous gauge boson couplings,
the subject of this meeting. As discussed by Wudka \cite{wudka},
there are several ways in which these anomalous gauge boson
couplings may be written in terms of a low energy effective
Lagrangian.

I choose to study the case of a strongly interacting symmetry
breaking sector in which there is no light Higgs boson, and
therefore, use an effective Lagrangian with a non-linearly
realized symmetry breaking. My motivation for this choice
is simple: if there is a light Higgs boson we will find it directly
and not through its contributions to anomalous couplings.
I furthermore choose the ``Gasser and Leutwyler''\cite{gale}
construction of the effective Lagrangian because it makes the
discussion of global symmetries transparent.

First I briefly review the formalism in order
to establish the notation and discuss the possible size of
the parity violating anomalous coupling from simple dimensional
analysis. I then study the indirect bounds  that can be placed on
this coupling from its one-loop contribution to rare decays and
partial $Z$ widths.
Finally I discuss how to isolate the parity violating coupling in
future high energy experiments.

\section{Formalism}

\subsection{Effective Lagrangian}

The starting point is the minimal standard model without a Higgs
boson. This model can be written as the usual standard model,
but replacing the scalar sector with the effective Lagrangian
\cite{longo}:
\begin{equation}
{\cal L}^{(2)}={v^2 \over 4}\sp{D^\mu \Sigma^\dagger D_\mu \Sigma}.
\label{lagt}
\end{equation}
The matrix $\Sigma \equiv \exp(i\vec{w}\cdot \vec{\tau} /v)$, contains the
would-be Goldstone bosons $w_i$ that give the $W$ and $Z$ their
mass via the Higgs mechanism. Their interactions with the $SU(2)_L
\times U(1)_Y$ gauge bosons follow from the covariant derivative:
\begin{equation}
D_\mu \Sigma = \partial_\mu \Sigma +{i \over 2}g W_\mu^i \tau^i
-{i \over 2}g^\prime B_\mu \Sigma \tau_3.
\label{covd}
\end{equation}

The details of the physics that break electroweak symmetry determine
the next-to-leading order effective Lagrangian. At energies small
compared to $\Lambda$, it is sufficient to consider those terms
that are suppressed by $E^2/\Lambda^2$ with respect to Eq.~\ref{lagt}.
There are three terms in this next to leading order effective Lagrangian
that contribute to gauge boson self-energies at tree level
(and thus to the LEP observables $\epsilon_{1,2,3}$ of Ref.\cite{alta}).
For later reference, the one the respects the custodial symmetry is
$L_{10}$.

There are several terms in the next to leading order effective Lagrangian
that contribute to three gauge boson couplings at tree level.
Only two of them respect the custodial symmetry, for later reference they are
$L_{9L}$, $L_{9R}$.

Finally, there are also several terms in the next to leading order effective
Lagrangian that contribute at tree level to couplings with at least
four gauge bosons.
Two of these terms respect the custodial symmetry and for
later reference they are $L_{1}$, $L_{2}$.

The next to leading order effective Lagrangian that respects the custodial
symmetry is then:
\begin{eqnarray}
{\cal L}^{(4)}\ &=&\ {v^2 \over \Lambda^2}  \biggl\{ L_1 \, \biggl[
\sp{D^\mu\Sigma^\dagger D_\mu \Sigma} \biggr]^2
\ +\  L_2 \,
 \sp{D_\mu\Sigma^\dagger D_\nu \Sigma}
\sp{D^\mu\Sigma^\dagger D^\nu \Sigma}   \nonumber \\
& -& i g L_{9L} \,\sp{W^{\mu \nu} D_\mu
\Sigma D_\nu \Sigma^\dagger}
\ -\ i g^{\prime} L_{9R} \,\sp{B^{\mu \nu}
D_\mu \Sigma^\dagger D_\nu\Sigma} \nonumber \\
& +& g g^{\prime} L_{10}\, \sp{\Sigma
B^{\mu \nu}
\Sigma^\dagger W_{\mu \nu}} \biggr\}.
\label{lfour}
\end{eqnarray}
There are many more terms that break the
custodial symmetry, but only one that violates parity while conserving
${\cal CP}$. This term gives rise to three and four gauge boson
couplings and is the subject of this talk.

The motivation for considering this term is, of course, that we should
explore all possibilities for the symmetry breaking sector.  In
theories where the electroweak symmetry breaking sector conserves parity,
like the minimal standard model or most technicolor theories, this term
is expected to be very small.

The parity violating and ${\cal CP}$ conserving effective Lagrangian at
order $1/\Lambda^2$ is
\begin{equation}
{\cal L}^{(4)}_{\rm p.v.}=\norm
g {\hat \alpha}\epsilon^{\alpha \beta
\mu \nu}\sp{\tau_3 \Sigma^\dagger D_\mu \Sigma}
\sp{W_{\alpha \beta} D_\nu \Sigma \Sigma^\dagger}
\label{leffpv}
\end{equation}
where $W_{\mu\nu}$ is the $SU(2)$ field strength tensor. In terms of
$W_\mu \equiv W^i_\mu \tau_i$, it is given by:
\begin{equation}
W_{\mu\nu}={1 \over 2}\biggl(\partial_\mu W_\nu -
\partial_\nu W_\mu + {i \over 2}g[W_\mu, W_\nu]\biggr).
\end{equation}
It is easy to see that this is the only term that violates parity and
conserves ${\cal CP}$ to order $1/\Lambda^2$.

In unitary gauge, the effects of the Lagrangian Eq.~\ref{leffpv},
are very simple. There is a three gauge boson interaction:
\begin{equation}
{\cal L}^{(3)} =-  {\hat{\alpha}g^3 v^2\over \Lambda^2 c_\theta}
\epsilon^{\alpha \beta \mu \nu}\biggl(W^-_\nu \partial_\alpha
W^+_\beta - W^+_\beta \partial_\alpha W^-_\nu\biggr)Z_\mu,
\label{agblt}
\end{equation}
which generates a $Z(q) \ra W^+(p^+) W^-(p^-)$ coupling.
In the notation of Ref.\cite{hagi} we have the
correspondence:
\begin{equation}
g_5^Z = \hat{\alpha}{g^2 \over c^2_\theta}{v^2\over \Lambda^2}
={4 M_Z^2 \over \Lambda^2} \hat{\alpha}.
\label{gfdef}
\end{equation}
There is also a four gauge boson interaction required by electromagnetic
gauge invariance:
\begin{equation}
{\cal L}^{(4)}=i {2\hat{\alpha}g^4 v^2 s_\theta \over \Lambda^2 c_\theta}
\epsilon^{\alpha \beta \mu \nu}W^-_\alpha W^+_\beta
Z_\mu A_\nu.
\label{agblf}
\end{equation}
This interaction contributes to the processes we discuss and must be
considered simultaneously with that of Eq.~\ref{agblt}. The Feynman
rules for this interaction were written down in Ref.~\cite{dv}.

\subsection{Natural size of $g_5^Z$ and Unitarity}

Within the minimal standard model, the operator Eq.~\ref{leffpv} is
generated at one-loop by the splitting between top-quark and bottom-quark
masses. In the limit $m_t \gg m_W$, and setting $m_b=0$ one finds \cite{dv}:
\begin{equation}
\biggl({v^2 \over \Lambda^2}
\hat{\alpha}\biggr)_{\rm top} = {N_c \over 128 \pi^2}
\biggl(1-{8\over 3}s^2_\theta \biggr)\approx   10^{-3}
\label{topc}
\end{equation}
For comparison, in this same limit one obtains $\norm L_1
\sim -3 \times 10^{-4}$
and $\norm L_2 \sim 6 \times 10^{-4}$ \cite{dvf}.
We see that in this limit (in which the custodial symmetry is violated
``maximally''), the parity violating coupling is of the same size as
other anomalous couplings. Of course, this limit is not allowed by
the size of $\Delta\rho$. Taking the scale $\Lambda$ to be a few TeV
($2$~TeV for definiteness), one expects that in theories where there
is no custodial symmetry and $\rho \approx 1$ accidentally, $\hat{\alpha}$
can be of order one. On the other hand, in theories with a custodial
symmetry, one expects $\hat{\alpha}$ to be at most as large as $\Delta\rho$.

Our effective Lagrangian formalism breaks down at some scale
$\Lambda \leq 3$~TeV, and this manifests itself in amplitudes that
grow with energy and violate unitarity at some scale related to
$\Lambda$. By studying the high energy behavior of longitudinal
vector boson scattering one finds that the effective Lagrangian
description breaks down between 1 and 2 TeV. For our numerical
estimates we will work with energies up to 2 TeV. We thus turn the
question around and ask how large can $\hat{\alpha}$ be so that
all scattering amplitudes remain below their unitarity bound at
energies up to 2 TeV. The answer is $|\hat{\alpha}|< 5$. This means that
the bounds that can be placed at high energy experiments on this
coupling will only be meaningful if they are better than
$|\hat{\alpha}|< 5$. For bounds placed at lower energy machines such as LEP,
the bad high energy behavior shows up in the need for counterterms to the
one-loop calculations. Since we do not know what those counterterms are,
the bounds obtained will be connected to ``naturalness'' assumptions.

\section{Present Bounds}

In this section we study the bounds that already exist on $g_5^Z$.
They follow from considering the one-loop effects of the operator
Eq.~\ref{leffpv} in the coupling of a $Z$ boson to fermions. These
observables do not single out the effects of $g_5^Z$, they are sensitive
to most of the anomalous couplings.

\subsection{Rare $K$- and $B$-meson decays}

These rare decays receive contributions from the parity violating
effective Lagrangian Eq.~\ref{leffpv} at the one-loop level. One-loop
amplitudes with one vertex from the ${\cal O}(1/\Lambda^2)$
effective Lagrangian
are ${\cal O}(1/\Lambda^4)$. A complete study thus requires the next to next to
leading order counterterms, as well as two loop contributions from the
leading order effective Lagrangian. It is clear that there are several
contributions to these decays that occur at the same order as the one-loop
contribution from $g_5^Z$ and that they could cancel:
we assume that they do not.

In unitary gauge, the $g_5^Z$ coupling affects this decays by modifying
the ``$Z$-penguin'' diagram as discussed in Ref.\cite{dv,he}.
The result is dominated by top-quark intermediate states and is
finite due to a GIM cancellation.
With $x_t=m_t^2/m_W^2$ and defining
\begin{equation}
W(x_t)\equiv {3 \over 4}x_t\biggl({1\over 1- x_t}+{x_t \log x_t \over
(1-x_t)^2}\biggr)
\label{wxt}
\end{equation}
one can write down the result in terms of the notation of Ref.~\cite{burasnew},
by replacing:
\begin{eqnarray}
Y(x_t) &\ra & \hat{Y}(x_t)=
Y(x_t) + g_5^Z c^2_\theta W(x_t) \nonumber\\
Y(x_t) &=& {x_t \over 8}\biggl({x_t-4 \over x_t -1}+{3x_t \over (x_t-1)^2}
\log x_t\biggr)
\label{repl}
\end{eqnarray}
One finds for example:
\begin{equation}
\Gamma (B_s\ra \mu^+ \mu^-)={G_F^2 \over \pi}\biggl({\alpha \over 4 \pi
s^2_\theta}\biggr)^2F^2_B m^2_\mu m_B|V_{tb}V^*_{ts}|^2 \hat{Y}(x_t)^2.
\label{bmumu}
\end{equation}
There are similar contributions to the decays $K_L \ra \mu^+ \mu^-$
and $K^+ \ra \pi^+ \nu \overline{\nu}$. Because $K_L \ra \mu^+ \mu^-$
is dominated by long distance physics, it can only be used to place a
``theoretical'' bound on $g_5^Z$ by requiring the new contribution to be less
than the standard model short distance contribution \cite{he}. This results
in
\begin{equation}
g_5^Z < {\cal O}(1).
\label{raredecay}
\end{equation}
If the rates for the short distance dominated
processes $B_s \ra \mu^+ \mu^-$ or
$K^+ \ra \pi^+ \nu \overline{\nu}$ are measured to within factors of
two, the same bound Eq.~\ref{raredecay} will be obtained. To improve this
bound would require a precision measurement of the rate, combined with
detailed knowledge of all the standard model parameters (CKM angles, top
quark mass, and decay constants) \cite{dv}.

\subsection{Partial $Z$ widths at LEP}

High precision measurements at the $Z$ pole at LEP combined with
polarized forward backward asymmetries at SLC put stringent
limits on any new physics beyond the standard
model. These measurements are now sufficiently precise to limit
the one-loop contribution of anomalous three gauge boson couplings to
the $Z$ pole observables.

The bounds arise because at the one-loop level Eq.~\ref{leffpv}
modifies the $Z f^\prime \overline{f}$ couplings. Because the operator
modifies the gauge boson self-couplings, its one-loop effects on
the $Z$ couplings to fermions affect both the flavor diagonal and the
flavor changing vertices considered in the previous section.
It turns out that the flavor diagonal vertices provide better
constraints due to the extraordinary precision of the LEP
measurements.

The flavor diagonal calculation is different from the flavor changing
calculation in that the one-loop effects of $g_5^Z$ are now divergent.
Nevertheless, from the effective field theory perspective this is not
significant. In both cases, there are other contributions to the physical
processes from other non-renormalizable interactions between fermions
and gauge bosons. In the previous section we adopted the point of view that
those other interactions did not cancel the contributions of $g_5^Z$ to
rare decays. In this section we adopt the point of view that the
renormalization of such couplings removes any divergence from physical
amplitudes. As is usual in effective field theory calculations, we
estimate the size of the $g_5^Z$ contribution to the physical amplitudes
from the leading non-analytic terms that go like $\log(\mu)$.

%We find  that $g_5^Z$ affects only the left handed
%coupling of the $Z$ to fermions and its effects can be incorporated
%by modifying the tree level coupling in the form \cite{dvlep}:
%$L_f\rightarrow L_f + \eta c^2_\theta /s^2_\theta$ where
%\begin{equation}
%\eta={3\alpha \over 2\pi}g_5^Z \log\biggl({\mu\over M_W}\biggr).
%\end{equation}

In addition to the direct contribution of $g_5^Z$ to the $Z f\overline{f}$
vertex we must consider indirect effects due to renormalization. In particular,
the operator of Eq.~\ref{lfour} also modifies the $W^\pm \rightarrow \ell^\pm
\nu$ coupling, contributing in this way to muon decay and thus introducing
a renormalization of $G_F$. In terms of the input parameters:
$G_F$ as measured in muon decay, $\alpha_*(M_Z^2)
\approx 1/128.8$ \cite{peskin}
and the physical $Z$ mass, and using
a $s^2_\theta$ defined by the relation:
\begin{equation}
s^2_Z c^2_Z \equiv {\pi \alpha_* \over \sqrt{2} G_F M^2_Z} ,
\end{equation}
we find:
\begin{equation}
{\delta\Gamma^{5}_{f} \over \Gamma^{(0)}_{f} }=
{3\alpha \over 2\pi}g_5^Z \log\biggl({\mu\over M_W}\biggr)
\biggl[{2 L_f \over L_f^2 +R_f^2} {c^2_\theta \over s^2_\theta}
+ \biggl(1+{2 R_f(L_f + R_f) \over
L_f^2 +R_f^2}{c^2_\theta \over s^2_\theta - c^2_\theta}\biggr)\biggr].
\label{modvertex}
\end{equation}
Where the shifts in the partial decay widths of the $Z$ are defined by:
\begin{equation}
\Gamma (Z \rightarrow f \overline{f}) = \Gamma^{SM}_{f} + \delta\Gamma^5_{f}
\equiv
\Gamma^{SM}_{f}\biggl(1 + {\delta\Gamma^5_{f} \over
\Gamma^{SM}_{f} } \biggr).
\end{equation}

To place bounds on $g_5^Z$ (and other couplings) we compare the standard model
predictions, $\Gamma_{f}^{SM}$, including the one loop QED and QCD
radiative corrections with the most recent results from LEP.
We use the theory numbers of Langacker \cite{smpred}.

Our $90\%$ confidence level interval for the
allowed values of $g_5^Z$ is shown in Table~\ref{t: pwid} \cite{dvlep}.
To place this result in perspective, it is instructive to compare them
with results for the couplings that respect the custodial symmetry in
Eq.~\ref{lfour} \cite{davalep}:
\begin{table}[htb]
\centering
\caption[]{$90\%$ confidence level intervals for $g_5^Z$ from different LEP
observables.}
\begin{tabular}{c c} \hline
Coupling ($\Lambda = 2$~TeV) & $90\%$ confidence level interval  \\ \hline
$L^r_{10}(M_Z)_{new}$  &  (-0.46,0.77) \\
$L_{9L}$ & (-22,16) \\
$L_{9R}$ & (-77,94) \\
$L_1+5/2 L_2$ & (-28,26) \\
$\hat{\alpha}$ & (-9,5) \\
$g_5^Z$ & (-0.07,0.04) \\ \hline
\end{tabular}
\label{t: pwid}
\end{table}

{}From Table~\ref{t: pwid} we see that the best limits are placed on the
coupling that contributes to the $Z$ self-energy at tree level, $L_{10}$.
The bounds on the other couplings are obtained by taking only one of them
to be non-zero at a time, and they are all comparable. A deviation in the
partial $Z$ widths from their standard model value could not be attributed to
a single coupling. In order to isolate the effects of $g_5^Z$ we consider in
the next two sections other observables that single out the parity violating
operator.

\section{Future Bounds}

In this section we discuss the most promising reactions to place bounds
on $g_5^Z$ in future colliders. These bounds arise from considering
observables that single out the coupling $g_5^Z$ making through its
parity violating nature.

\subsection{Forward-backward asymmetry in $e^+_L e^-_R \ra W^+ W^-$}

In this section we study the effect of the parity violating operator
Eq.~\ref{leffpv} on the process $e^+ e^- \ra W^+ W^-$. This process
receives contributions from $s$-channel $\gamma$ and $Z$ exchange diagrams
and from a $t$-channel neutrino exchange diagram. The latter contributes
only to $e^-_L e^+_R \ra W^- W^+$.

The differential cross-section for right-handed electrons is found to be
\cite{dv,ahn}:
\begin{eqnarray}
{d\sigma_{TT} \over d(\cos\theta)} \biggr|_{e^-_R}
&=& {\pi \alpha^2 \over  s} \beta^3
{m^4_Z \over (s-m^2_Z)^2}\sin^2\theta \nonumber \\
{d\sigma_{LL} \over d(\cos\theta)} \biggr|_{e^-_R}
&=& {\pi \alpha^2 \over 32 s}
{\beta^3 \over c^4_\theta} {s^2 \over (s-m^2_Z)^2}
(5+\beta^2)^2\sin^2\theta \nonumber \\
{d\sigma_{TL} \over d(\cos\theta)} \biggr|_{e^-_R}
&=& {\pi \alpha^2 \over  s}
{\beta^3 \over c^2_\theta}
{m^2_Z s\over (s-m^2_Z)^2}\biggl(1+\cos^2\theta + 2\beta{s\over m^2_Z}g^Z_5
\cos\theta\biggr)
\label{diffc}
\end{eqnarray}
where $\beta^2=1-4m^2_W/s$. Other anomalous
couplings do not contribute to the forward backward asymmetry
in $e^-_R e^+_L \ra W^- W^+$ and they are not considered here.

As can be seen from Eq.~\ref{diffc}, there is a term in $\sigma_{TL}$ that is
linear in $\cos\theta$ (the scattering angle in the center of mass). This
term gives rise to a
forward-backward asymmetry. Although there is a similar term in the
differential cross-section for $e^-_L e^+_R \ra W^+ W^-$, in that case one
also has a $t$-channel neutrino exchange diagram that gives rise to a very
large forward-backward asymmetry within the minimal standard model. Thus,
if we want to isolate the $g^Z_5$ term, it is very important to have
right-handed electrons. Since the cross-section for left-handed electrons
is several orders of magnitude larger than that for right-handed electrons,
it presents a formidable background.

\begin{figure}[h]
\centerline{\epsfysize=3in\epsfbox{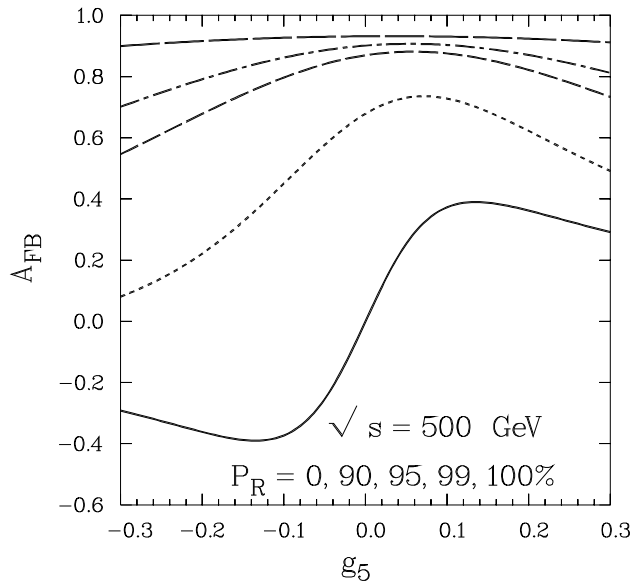}}
\caption[]{Forward-backward asymmetry for the process $e^+ e^- \ra W^+ W^-$ for
$\sqrt{s}=500$~GeV.
The different curves from upper most to lowest correspond to a fraction
of right handed electrons in the beam of 0\%, 90\%, 95\%, 99\% and 100\%.}
\label{fbasym}
\end{figure}

We find \cite{dv} that the largest sensitivity to $g_5^Z$ occurs in
the forward-backward asymmetry at high center of mass energies. This
sensitivity decreases dramatically if there is any contamination of
left handed electrons as shown in Figure~\ref{fbasym}, where we we
present the forward-backward asymmetry for an $e^+ e^-$ collider with
$\sqrt{s}=500$~GeV.
This figure shows the great sensitivity of the observable to the coupling
$g_5^Z$. Unfortunately it also shows how this sensitivity is lost if there
is even a small fraction of left handed electrons.

The total cross section is also sensitive to the value of $g_5^Z$, however,
a deviation in the total cross section from the standard model value would
not single out the $g_5^Z$ coupling.

\subsection{High energy $e^- \gamma \ra \nu W^- Z$}

In this section we explore the possibility of observing the effects of
the parity violating operator Eq.~\ref{leffpv} via the anomalous
four-gauge-boson coupling that it generates. We thus turn our attention
to high energy vector-boson fusion experiments. Given the
form of the four vector-boson interaction, Eq.~\ref{agblf}, we look at
processes involving one photon and one $Z$. There are several possibilities,
for example $Z\gamma$ production in high energy $e^+ e^-$ or $pp$ colliders.
This process, however, suffers from large standard model backgrounds.
We consider instead a high energy $e^- \gamma$ collider where we can
cleanly identify the process $e^- \gamma \ra \nu W^- Z$, and where we
can also consider a polarized photon if need be.

To understand the physics, we first use the equivalence theorem to
compute the the polarized cross sections for $W\gamma \ra wz$,
$\sigma (\lambda^W,\lambda^{\gamma})$ \cite{dv}:
\begin{eqnarray}
\sigma_{+-} &=\sigma_{-+}=&
{\pi\alpha^2 \over s^2_\theta}{1 \over 3s}
 \nonumber\\
\sigma_{++} &=\sigma_{--}=&
{\pi\alpha^2 \over s^2_\theta}{1 \over 3s}\biggl(
|g_5^Z|^2 c_\theta^4 {s^2\over m_W^4}\biggr) \nonumber\\
\sigma_{L+}&=\sigma_{L-}=&
{\pi\alpha^2 \over s^2_\theta}{1 \over 3s}\biggl(
|g_5^Z|^2 {c_\theta^4 \over 4}{s^3\over m_W^6}\biggr)
\label{swgpp}
\end{eqnarray}

{}From Eq.~\ref{swgpp} we see that the $g_5^Z$ term does not interfere with
the lowest order term. This
means that we can only construct observables sensitive to $g_5^Z$ that are
parity even and can thus be generated by other anomalous couplings.
However, it is possible that the cross section is
more sensitive to the $|g_5^Z|^2$ term than to those
terms proportional to $L_{9L}$, $L_{9R}$ or $L_{10}$ in very high energy
machines. The reason is that the $|g_5^Z|^2$ term is
the only one that contributes to the amplitude where all three vector-bosons
are longitudinally polarized (this is the source of $\sigma_{L\pm}$ in
Eq.~\ref{swgpp}) and we expect these terms of ``enhanced electroweak
strength'' to dominate at high energies. This is indeed the case, as
shown by a numerical simulation \cite{cdhv}.

To demonstrate the significant sensitivity of this process to $\hat \alpha$,
we consider a 2 TeV $e^+e^-$ collider (the $e^- \gamma$
differential cross section is folded with the energy spectrum of the
back scattered photon). We make use of the relative enhancement
of $\hat{\alpha}$ at higher energies to isolate this coupling with a
set of cuts like:
\begin{equation}
|\cos\theta_V |< 0.8, \quad p_T^{}(WZ) > 30~{\rm GeV}, \quad
M(WZ) > 0.5~{\rm TeV}.
\label{cutsiii}
\end{equation}
Here the $p_T(WZ)$  cut  is optimized to suppress reducible backgrounds
from other sources.
The numerical results support our earlier conclusion. The interference of
$g_5^Z$ with the lowest order term (which vanishes in the effective $W$
approximation) is very small, so it is not possible to single out the $g_5^Z$
term through a parity violating observable. On the other hand, by isolating
the high invariant mass region for the $WZ$ pair, we significantly
enhance the contribution of $g_5^Z$ with respect to other couplings as
shown in Figure~2. In that Figure we show a $3 \sigma$ significance
resulting from the anomalous couplings
 $\hat \alpha, L_{9L}$, and $L_{9R}$,
at $\sqrt{s_{ee}}=2$~TeV
for  the cuts of Eq.~\ref{cutsiii},
as a function of the integrated luminosity.
We see that the coefficient
$\hat \alpha$ can be probed  here to a level less than $1~(\Lambda /2~{\rm
TeV})^2$.
\begin{figure}[htp]
\vspace{-2.5in}
\centerline{\epsfxsize=4.5in\epsfbox{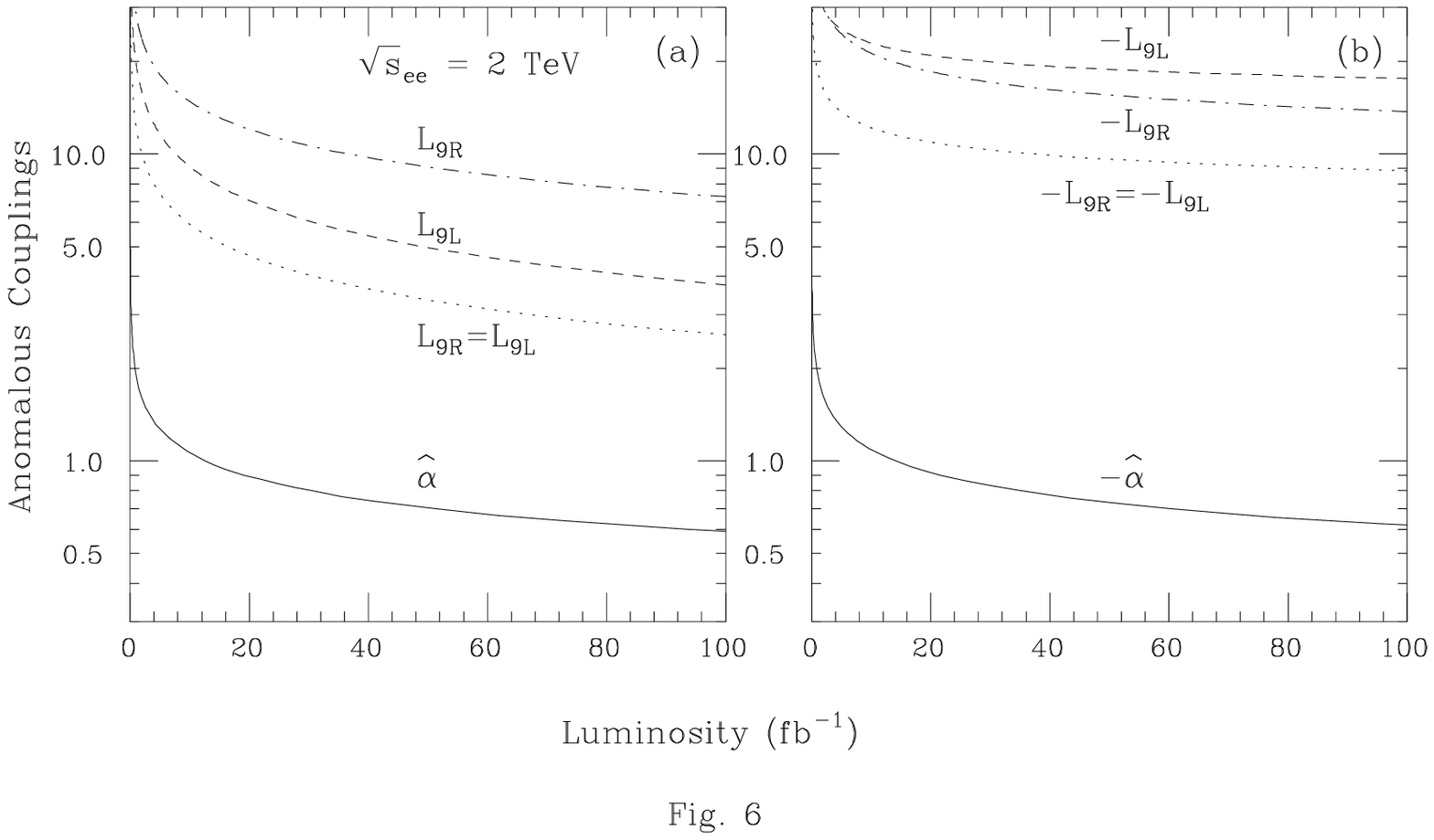}}
%\vspace{3.7in}
\vspace{-1.2in}
\caption[]{$3\sigma$ sensitivity of an $e^+ e^-$ collider at $\sqrt{s_{ee}}=2$
TeV (operating in the $e^- \gamma$ mode) to $\hat{\alpha},  L_{9L}$
and $L_{9R}$ with the cuts Eq.~\ref{cutsiii}.
The curves are  shown as a function of integrated luminosity, and we
set $\Lambda = 2$~TeV.}
\label{egamma}
\end{figure}

\section{Conclusions}

Here we summarize the bounds that can be placed on the coupling $g_5^Z$
from all the processes discussed in this talk. We also compare them with
the natural size expected for $g_5^Z$. We see from Table~2 that the
current bound at LEP is an order of magnitude better than the bounds
from rare decays.
An indication of how precise the LEP measurements are is the fact that
the LEP bound can only be improved by one order
of magnitude in a 2~TeV $e^+e^-$ collider. In principle,
$g_5^Z$ can be bound precisely by studying the forward backward
asymmetry in $e^+_L e^-_R \ra W^+ W^-$, however, it is not clear that it
will ever be possible to achieve the high degree of polarization that
would be required.
\begin{table}[h]
\centering
\caption[]{Comparison of Bounds on $g_5^Z$.}
\begin{tabular}{l l}\hline
Process & Bound on $|g_5^Z|$  \\ \hline
Rare Decays & ${\cal O}(1)$ \\
Partial $Z$ widths & $5 \times10^{-2}$ \\
$A_{fb}(e^+_L e^-_R \ra W^+ W^-)$ & Potentially very good, needs
$P\sim 100\%$ \\
$e^- \gamma \ra \nu W^- Z$  &  $5 \times10^{-3}$
(in a 2 TeV $e^+e^-$ collider) \\
Natural size & $10^{-4}$ (with custodial symmetry) \\
&              $10^{-2}$ (without custodial symmetry) \\ \hline
\end{tabular}
\label{t: conc}
\end{table}
{}From the numbers in Table~2 we conclude that an
observation of a non-zero value for $g_5^Z$ would be  very strong
evidence against a custodial symmetry in the electroweak symmetry
breaking sector.

\section{Acknowledgements}

This work was supported in part by a DOE OJI award under contract
number DEFG0292ER40730. I am grateful to S. Dawson for many pleasant
collaborations on the matters of this talk. I also thank
my collaborators T. Han and K. Cheung.

\end{document}